\documentclass[]{article}

\usepackage{graphicx}

\def\uminus{\left. u\right|_-}
\def\uplus{\left. u\right|_+}

\begin{document}

\title{Remarks on some open problems in phase-field modelling of solidification}

\author{Mathis Plapp\\
Physique de la Mati\`ere Condens\'ee,\\ \'Ecole Polytechnique, CNRS,
91128 Palaiseau, France}

\maketitle

\begin{abstract}
Three different topics in phase-field modelling of solidification 
are discussed, with particular emphasis on the limitations of the 
currently available modelling approaches. First, thin-interface
limits of two-sided phase-field models are examined, and it is
shown that the antitrapping current is in general not sufficient to
remove all thin-interface effects. Second, orientation-field models
for polycrystalline solidification are analysed, and it is shown
that the standard relaxational equation of motion for the orientation
field is incorrect in coherent polycrystalline matter. Third, it is
pointed out that the standard procedure to incorporate fluctuations 
into the phase-field approach cannot be used in a straightforward way 
for a quantitative description of nucleation.  
\bigskip

\end{abstract}

\section{Introduction}

The phase-field method has become the method of choice for simulating
microstructure formation during solidification. It owes its popularity
mainly to its algorithmic simplicity: the cumbersome problem of tracking
moving solid-liquid interfaces or grain boundaries is avoided by describing 
the geometry in terms of one or several phase fields. The phase fields
obey simple partial differential equations that can be easily coded by
standard numerical methods. 

The foundations of the phase-field method and its application to
solidification have been the subject of several recent review articles 
\cite{Boettinger02,Chen02,Granasy04,Plapp07,Singer08,Emmerich08,Steinbach09},
and it seems of little use to repeat similar information here. Instead, 
in this paper several topics are discussed where robust phase-field 
modelling tools are not yet available because some fundamental 
questions remain open. In Sec.~\ref{sec2}, the thin-interface limit 
of two-sided phase-field models is examined, and it is shown that the 
currently available approaches cannot in general eliminate all effects linked
to the finite interface thickness. In Sec.~\ref{sec3}, orientation-field
models for polycrystalline solidification are discussed, and it
is shown that the standard equation of motion usually written down 
for the orientation field is not appropriate for the evolution of
coherent crystalline matter. Finally, in Sec.~\ref{sec4}, the
inclusion of microscopic fluctuations in the phase-field equations
is reviewed, and it is shown that the standard approach cannot be 
used in a straightforward way to investigate the process of nucleation.

The common point of these topics is that they pose challenges or 
limitations for straightforward computations. Indeed, a characteristic
feature of the phase-field method is that its equations can often be
written down following simple rules or intuition, but that their 
detailed properties (which have to be known if quantitative simulations 
are desired) become only apparent through a mathematical analysis
that can be quite involved.
Therefore, it is not always easy to perceive the limits of applicability
of the method. It is hoped that the present contribution will be
helpful to point out some pitfalls and to stimulate further discussions
that will facilitate the solution of these issues.

\section{Thin-interface limits: antitrapping current and the Kapitza resistance}
\label{sec2}

The precision and performance of phase-field models have been
greatly enhanced in the last decade by a detailed control of their 
properties. Phase-field models are rooted in the mean-field 
description of spatially diffuse interfaces by order parameters.
However, to be useful for simulating microstructure formation
in solidification, phase-field models need to bridge the scale 
gap between the thickness of the physical solid-liquid interfaces 
and the typical scale of the 
microstructures. This is achieved by increasing the interface
width in the model, sometimes by several orders of magnitude.
Obviously, this procedure magnifies any physical effect that is
due to the diffuseness of the interface. Therefore, to guarantee
precise simulations, all these effects have to be
controlled and, if possible, eliminated. The privileged tool
to achieve this is the so-called {\em thin-interface limit}:
the equations of the phase-field model are analysed under the
assumption that the interface thickness is much smaller than any
other physical length scale present in the problem, but otherwise arbitrary.
The procedure of matched asymptotic expansions then yields the
effective boundary conditions valid at the macroscale, which
contain all effects of the finite interface thickness up to the
order to which the expansions are carried out.

This procedure was pioneered by Karma and Rappel, who analysed
the symmetric model of solidification (equal diffusion constants
in the solid and the liquid) and obtained a thin-interface correction 
to the expression of the kinetic coefficient \cite{Karma96RC}. The use of 
this result has made it possible to carry out quantitative simulations 
of free dendritic growth of a pure substance, both at high and low
undercoolings \cite{Karma98,Provatas99,Karma00,Bragard02}. It turned out,
however, that the generalisation of this method to a model with
arbitrary diffusivities is far from trivial \cite{Almgren99}, since
several new thin-interface effects appear, which cannot all be
eliminated simultaneously. A solution to this problem was found
later for the case of the one-sided model (zero diffusivity in
the solid) with the introduction of the so-called antitrapping
current \cite{Karma01}, and it was shown that quantitative simulations
of alloy solidification are possible with this model \cite{Echebarria04},
including multi-phase \cite{Folch03,Folch05} and 
multi-component alloys \cite{Kim07}.
Recently, several extensions of the antitrapping current were
put forward to generalise the approach to the case of finite 
diffusivity in the solid \cite{Steinbach09,Gopinath06,Ohno09,Ducousso09}, 
and simulations were presented which show that 
the approach works well for the instability of a steady-state
planar interface \cite{Gopinath06} and for free dendritic growth \cite{Ohno09}.
However, as will be shown below, this is only a partial solution to 
the problem of developing a general quantitative model, since 
there is a second, independent thin-interface effect that cannot 
be removed by an antitrapping current, namely, the Kapitza resistance. 

For the sake of concreteness, consider the standard phase-field
model for the solidification of a pure substance as discussed 
in Refs.~\cite{Karma98,Almgren99}. The evolution equation for the
phase field reads
\begin{equation}
\tau \partial_t \phi = W^2{\vec \nabla}^2 \phi + \phi-\phi^3 -\lambda u 
(1-\phi^2)^2,
\label{pf}
\end{equation}
where $\phi$ is the phase field, with $\phi=1$ and $\phi=-1$ corresponding
to solid and liquid, respectively, $\tau$ is the relaxation time of the
phase field, $W$ is the interface thickness, and $\lambda$ is a
dimensionless coupling constant. The field $u$ is a dimensionless
temperature defined by $u=(T-T_m)/(L/c_p)$, where $T_m$, $L$
and $c_p$ are the melting temperature, latent heat,
and specific heat, respectively. It is assumed for simplicity
that $c_p$ is the same in both phases. The temperature is governed
by a diffusion equation with a source term,
\begin{equation}
\partial_t u = \vec\nabla \left[D(\phi)\vec\nabla u\right]
   + \frac 12 \partial_t h(\phi).
\label{diffusion}
\end{equation}
Here, $h(\phi)$, which satisfies $h(\pm 1)=\pm 1$, is a function that 
describes the release or consumption of latent heat during the phase 
transition, and $D(\phi)$ interpolates between the thermal diffusivities
of the liquid and the solid, $D_l$ and $D_s$,
\begin{equation}
D(\phi) = D_l q(\phi),
\label{ddef}
\end{equation}
where the interpolation function $q(\phi)$ satisfies $q(1)=D_s/D_l$
and $q(-1)=1$. For simplicity, crystalline anisotropy 
has not been included in the above model because it is not 
necessary for the present discussion. Furthermore, the equations
have been stated in the language of a two-sided thermal model, but 
with some modifications (as detailed in Refs.~\cite{Echebarria04,Ohno09}), 
they also apply to the isothermal solidification of a binary alloy. 
In this case, $u$ is a dimensionless chemical potential (conjugate 
to the concentration of one of the alloy components), and $D(\phi)$ 
is the chemical diffusivity.

In the following, two simple one-dimensional solutions of these
equations will be analysed. The first is a steady-state planar
front that propagates with constant velocity $V$ in the positive
$x$ direction into a liquid of undercooling $\Delta$ ($u\to -\Delta$ 
for $x\to\infty$), and leaves behind a constant temperature. 
This solution only exists if the liquid is undercooled beyond 
the hypercooling limit, that is, $\Delta >1$. The sharp-interface
solution to this problem is readily obtained and reads
\begin{eqnarray}
u & = {\rm const.} = \uminus \qquad\qquad\qquad\qquad 
    \qquad & \mbox{\rm in the solid}\;\;(x<0) \\
u & =  -\Delta + \left( \uplus + \Delta\right) \exp(-xV/D_l)
    \qquad & \mbox{\rm in the liquid}\;\;(x>0)
\end{eqnarray}
for an interface located at $x=0$ (in the frame moving with the 
interface). Here, $\uminus$ and $\uplus$ are the limit values
of the temperature when the interface is approached from the solid
and the liquid side, respectively. In the standard formulation
of the free boundary problem of solidification, it is assumed
that the temperature is the same on the two sides of the interface,
$\uminus=\uplus$. Then, the use of the two boundary conditions
$\uplus = -\beta V$, where $\beta$ is the linear kinetic coefficient,
and $V=-D_l\left.\partial_xu\right|_+$ (the Stefan boundary condition)
determines the solution, $\uplus=\uminus=-\Delta+1$ (a simple 
consequence of heat conservation), and $V=(\Delta - 1)/\beta$.

The phase-field equations can be analysed and related to this
sharp-interface solution by the method of matched asymptotic 
expansions in the limit where the interface thickness $W$ is 
much smaller than the diffusion length $D_l/V$. This calculation 
has been presented in detail in 
Refs.~\cite{Karma98,Almgren99,Echebarria04,Ohno09} and will
not be repeated here. The essential outcome is that, in general,
the two asymptotes of the bulk phases do {\em not} correspond
to the same temperature. The difference is given, to the lowest
order, by
\begin{equation}
\uminus - \uplus = \frac{V}{2} 
   \left[\int^{-\infty}_0 \frac{h(\phi_0)-1}{D(\phi_0)}\; dx
        - \int_0^\infty\left(\frac{h(\phi_0)-1}{D(\phi_0)}+
          \frac{2}{D_l}\right)\;dx\right],
\label{tjump1}
\end{equation}
where $\phi_0(x)$ is the equilibrium profile of the phase
field. The physical interpretation of this temperature jump is
{\em trapping}: when the diffusivity decreases upon solidification, 
the heat generated at the rear of the interface gets trapped. 
In the alloy version of the model, this is nothing but the well-known 
solute trapping effect. Indeed, in sharp-interface models of
alloy solidification the chemical potential exhibits a jump 
at the interface when solute trapping occurs. In the 
phase-field model, the temperature profile through the interface 
is determined by the interplay between
the rejection of latent heat and the diffusion away from
the interface; therefore, it is natural that the heat
source function $h(\phi)$ and the diffusivity function
$D(\phi)$ appear in Eq.~(\ref{tjump1}).

Whereas, thus, this discontinuity is physically correct,
it generates problems for simulations. To see this,
is is sufficient to rewrite Eq.~(\ref{tjump1}) in order
to make the relevant scales apparent. Since the only length
scale in Eq.~(\ref{pf}) is the interface thickness $W$, the
equilibrium solution $\phi_0$ is a function only of the
reduced variable $\eta=x/W$. Using this together with the
interpolation of $D(\phi)$ given by Eq.~(\ref{ddef}), 
Eq.~(\ref{tjump1}) becomes
\begin{equation}
\uminus - \uplus = \frac{VW}{2D_l} \left(F_--F_+\right),\qquad {\rm with}
\label{tjump1a}
\end{equation}
\begin{equation}
F_\pm = \int_0^{\pm\infty} 
   \left[ p(\phi_0(\eta))-p(\pm 1)\right]\; d\eta\qquad {\rm and}
\label{fdef}
\end{equation}
\begin{equation}
p(\phi) = \frac{h(\phi)-1}{q(\phi)}.
\label{pdef}
\end{equation}
The temperature jump is thus proportional to the velocity,
the interface thickness, and the difference of the two integrals;
the latter depends only on the choice of the interpolation functions. 
If $W$ is the physical interface thickness (a few Angstroms), this effect
is negligibly small, but if the interface thickness is increased
by a large factor to make simulations feasible, this leads to 
potentially large errors in the simulations.

As discussed in detail in Refs.~\cite{Almgren99,Karma01,Echebarria04},
it is not possible to eliminate this macroscopic discontinuity
simply by the choice of appropriate interpolation functions,
due to other constraints not discussed here. The solution put
forward in Ref.~\cite{Karma01} and further developed in 
Ref.~\cite{Echebarria04} is the introduction of an antitrapping
current: Eq.~(\ref{diffusion}) is replaced by
\begin{equation}
\partial_t u = \vec\nabla \left(D(\phi)\vec\nabla u - \vec j_{at}\right),
\end{equation}
where the antitrapping current $\vec j_{at}$ is given by
\begin{equation}
\vec j_{at} = a(\phi) W \dot\phi \hat n,
\end{equation}
where $\dot\phi$ is a shorthand for the time derivative $\partial_t\phi$,
$\hat n = -\vec\nabla \phi /|\vec\nabla \phi|$ is the unit
normal vector to the interface, and $a(\phi)$ is a new interpolation
function. This term induces a current which is directed from the
solid to the liquid, and proportional to the interface velocity
(through the factor $\dot\phi$). It thus ``pushes'' heat from
the solid to the liquid side of the interface when the interface
moves, and can be used to adjust the temperature jump at the
interface. 
For the one-sided model ($D_s=0$) with the standard choices
$h(\phi)=\phi$ and $q(\phi)=(1-\phi)/2$, it was shown that
a constant $a(\phi)\equiv 1/(2\sqrt{2})$ leads to a vanishing
jump in $u$, because it modifies the function $p(\phi)$ in
Eq.~(\ref{fdef}) such that $F_+=F_-$. Thus, continuity of the temperature 
between the two sides of the interface (local equilibrium) is restored 
for arbitrary $W$ and $V$, as long as the asymptotic analysis remains valid.

Recently, several authors have put forward generalisations of
this approach \cite{Steinbach09,Ohno09,Ducousso09}
for arbitrary ratio of the diffusivities. For the case analysed
above (that is, the current far inside the solid vanishes), 
they reduce to the simple prescription that the same expression
for the antitrapping current can be used, but with an additional
prefactor that can be written as $(1-D_s/D_l)$,
\begin{equation}
\vec j_{at} = a\left(1-\frac{D_s}{D_l}\right) W \dot \phi \hat n.
\label{antinew}
\end{equation}
Indeed, the asymptotic analysis shows \cite{Gopinath06,Ohno09,Ducousso09} 
that in this way the temperature jump can be eliminated.

However, this is not the only thin-interface effect that can
arise in the two-sided case. To see this, consider now a different
situation, namely an immobile interface in a temperature
gradient. Such an interface can be easily obtained in experiments
by maintaining a pure substance between two walls which are held below 
and above the melting temperature, respectively. When the interface
is stationary, $\partial_t\phi=\partial_tu=0$ by definition, and 
Eq.~(\ref{diffusion}) implies that the system is crossed by a constant 
heat current flowing from the liquid into the solid,
\begin{equation}
-D(\phi)\partial_x u = -j,
\label{current}
\end{equation}
with $j$ a positive constant. As before, the centre of the
interface is located at $x=0$, and the solid is located in the
domain $x<0$. This situation can be analysed without performing
a perturbation expansion, since it is sufficient to integrate
Eq.~(\ref{current}) to obtain a solution for $u$,
\begin{equation}
u(x) = \bar u + \int_0^x \frac{j}{D(\phi(x))}\;dx,
\end{equation}
where $\bar u$ is the temperature at $x=0$. The sharp-interface 
solution for this case is simply given by
\begin{eqnarray}
u(x) & = \uminus + (j/D_s)\; x \quad & \mbox{\rm in the solid} \\
u(x) & = \uplus + (j/D_l)\; x \quad & \mbox{\rm in the liquid}.
\end{eqnarray}
Matching the asymptotes of the phase-field and sharp-interface 
expressions, it is straightforward to show that there is again 
a temperature jump given by
\begin{equation}
\uplus - \uminus =j\left[ 
   \int_0^\infty \left(\frac{1}{D(\phi(x))}-\frac{1}{D_l}\right)\; dx
 - \int_0^{-\infty} \left(\frac{1}{D(\phi(x))}-\frac{1}{D_s}\right)\;dx \right],
\end{equation}
this time proportional to the {\em current}. If the phase-field
profile is replaced by its equilibrium shape, this can be rewritten as
\begin{equation}
\uplus - \uminus = \frac{jW}{D_l} \left(G_+-G_-\right)
\label{tjump2}
\end{equation}
with
\begin{equation}
G_\pm = \int_0^{\pm\infty} \left(\frac{1}{q(\phi_0(\eta))}
   - \frac{1}{q(\pm 1)}\right) \; d\eta.
\end{equation}

This temperature jump corresponds to a surface thermal resistance, 
also called Kapitza resistance, first found for an interface between 
liquid helium and metal \cite{Kapitza41}. Indeed, in a sharp-interface
picture it is generally necessary to assign a surface resistance to 
an interface for a complete description of heat transfer, because
transport through an interface can be decomposed into three elementary
steps: transport in one bulk phase, crossing of the interface, and 
transport in the other phase. The surface resistance describes the
kinetics associated with the crossing of the interface (its inverse
is sometimes referred to as the interfacial transfer coefficient).
It is characterised either by the value of the resistance, 
$(\uplus - \uminus)/j$, or by a length that is obtained by dividing 
this resistance by the conductivity of the liquid phase. 
Here, this characteristic length is simply $W(G_+-G_-)$, which 
is of the order of the interface thickness. Since this quantity is
actually an interface excess of the inverse diffusivity (in complete
analogy to the interface excesses for equilibrium quantities
obtained by the well-known Gibbs construction), it can also be 
negative -- this does not violate the laws of thermodynamics because
the {\em local} transport coefficients are strictly positive. If the surface 
resistance is finite, the temperature in the sharp-interface model is {\em not} 
continuous at the interface, but exhibits a jump that is proportional to
the current crossing the interface. In the alloy version of the model, 
this corresponds to a jump in chemical potential that is proportional 
to the solute flux \cite{Maugis94}. Such discontinuities have been 
thoroughly investigated \cite{Swartz89}, and can be measured 
in experiments \cite{Wolf83}
and detected in molecular dynamics simulations \cite{Barrat03,Xue03}
for solid-liquid interfaces.

Thus, like the trapping effect, the surface resistance is a
natural effect that is proportional to the interface thickness. 
If the interface thickness is to be upscaled, it should therefore 
also be eliminated. However, is is immediately clear that this 
effect cannot be eliminated by any antitrapping current proportional
to $\dot\phi$ as given by Eq.~(\ref{antinew}): since the interface 
does not move, $\dot \phi=0$ and the antitrapping current vanishes, 
independently of the current $j$ that crosses the interface.

The authors of both Refs.~\cite{Ohno09,Ducousso09} have recognised
the importance of the current $j$. They have developed generalised
expressions for the antitrapping current with coefficients that
depend on the value of $j$. As long as the interface velocity
remains non-zero, the formal asymptotic analysis shows that it
is still possible to eliminate the temperature jump. However,
for a fixed current $j$, the expressions of the coefficients
diverge when $V$ tends to zero, such that the asymptotic analysis
is not valid in this limit. Thus, it seems unlikely that this 
approach can be used as a robust method for simulations.

In summary, there exist two independent thin-interface effects, 
one proportional to $V$, and one proportional to $j$. On a very 
fundamental level, this is just the consequence of the fact
that the interface motion is driven by a diffusion equation, which
has two independent boundary conditions. The corresponding physical
quantities are the currents on the two sides of the interface, 
or one current and the velocity. A general solution to eliminate 
both thin-interface effects (which are linearly independent)
does not seem to exist at this moment, but the above considerations
can at least be used to obtain simple criteria when the
prescription of Eq.~(\ref{antinew}) can be used. Indeed, 
Eqs.~(\ref{tjump1a}) and (\ref{tjump2}) show that if $j\ll V$
(note that, since $u$ is dimensionless, $j$ has the dimension
of a velocity), the Kapitza effect is much smaller than the
trapping effect, and can thus be neglected. This is generally
the case for equiaxed dendritic growth, in which the gradients 
outside the growing dendrite, which determine the growth speed,
are much larger than the gradients inside the solid. Indeed,
it was shown in Ref.~\cite{Ohno09} that Eq.~(\ref{antinew})
works well in this case. However, problems might arise in
the case of alloy solidification in a temperature gradient
or for multicomponent alloys with widely different solute
diffusivities, since in this case large currents of heat
or certain solutes may cross an interface whose velocity is
controlled by a different diffusion field. Such cases have
to be critically examined before simulation results can be 
trusted.

\section{Polycrystalline solidification}
\label{sec3}

The size and shape of the crystalline grains formed upon solidification
is one of the most important factors that determine materials properties.
Therefore, phase-field models that are to be helpful for materials design
must be capable of dealing with the evolution of polycrystals, both during
solidification of individual columnar or equiaxed grains from the melt and
during the subsequent evolution of the grain structure after impingement.
This can be achieved using the multi-phase-field 
approach \cite{Steinbach96,Fan97,Garcke99,Steinbach99,Eiken06,Moelans08}, in 
which each grain is represented by a different phase field, even if they are of 
the same thermodynamic phase. The properties of each individual grain boundary 
or interface can then be specified separately \cite{Steinbach99}, and it has 
been demonstrated that good quantitative control of the grain boundary properties 
can be achieved \cite{Moelans08}. The problem of handling several hundreds or 
even thousands of
phase fields simultaneously can be solved by recognising that only a
small number of fields are important at any given point of 
space (see for example \cite{Vanherpe07}).

An alternative approach is the orientation-field method. Its starting
point is the remark that it would be desirable, both for efficiency and
simplicity, to formulate a model that works only with a small number of
field variables. Indeed, the orientation of a crystal can be described
by one scalar quantity (an angle) in two dimensions, and three scalars
in three dimension (for instance, the Euler angles). Orientation-field
models for pure substances in two dimensions that work with a single phase 
field, an orientation field (the local angle of the crystalline structure 
with respect to a fixed coordinate system), and the temperature field were
put forward in Refs.~\cite{Kobayashi00,Warren03}, and generalised for 
alloy solidification \cite{Granasy04} and to three 
dimensions \cite{Pusztai05,Kobayashi05}.
While these models are elegant and simple in their formulation and
therefore hugely appealing, it is pointed out here that the evolution
equation of the angle field, which takes the form of a simple relaxation
equation, does not correctly describe the microscopic evolution of the 
orientation field since it does not take into account the connectivity
of matter and the resulting geometrical conservation laws.

For simplicity, anisotropy and crystallographic effects will again
be neglected, and it is sufficient to consider a two-dimensional
system. The dimensionless free energy of the orientation-field 
model is \cite{Warren03}
\begin{equation}
{\cal F} =  \int \left[ \frac{W^2}{2}\left(\vec\nabla\phi\right)^2
  + s\tilde g(\phi)\left|\vec\nabla\theta\right|
  + \frac{\epsilon^2}{2}\tilde h(\phi)\left(\vec\nabla\theta\right)^2
  + f(\phi,u)\right] d\vec r,
\label{modelI}
\end{equation}
where now $\phi=0$ and $\phi=1$ in the liquid and the solid,
respectively, $s$ and $\epsilon$ are positive constants,
$\tilde g(\phi)$ and $\tilde h(\phi)$ are monotonous functions 
that satisfy $\tilde g(0)=\tilde h(0)=0$ and 
$\tilde g(1)=\tilde h(1)=1$, and $f(\phi,u)$
is the local free energy density, with $u$ the same dimensionless
temperature field as previously; the standard choice is
$f(\phi,u)=\phi^2(1-\phi)^2 +\lambda u (10\phi^3-15\phi^4+6\phi^5)$. 
Recently, an alternative model was developed \cite{Mellenthin07},
\begin{equation}
{\cal F} = \int \left[ \frac{W^2}{2}\left(\vec\nabla\phi\right)^2
  + \nu \frac{7\phi^3-6\phi^4}{(1-\phi)^2}\left(\vec\nabla\theta\right)^2
  + f(\phi,u) \right] d\vec r,
\label{modelII}
\end{equation}
where $\nu$ is a constant.
In the following, these models will be called model I and model II. 
They both have some features that distinguish them from standard 
phase-field models. Model I contains a term proportional to 
$|\vec\nabla\theta|$, which has a singular derivative at 
$|\vec\nabla\theta|=0$. Model II has only a regular square gradient 
term in $\vec\nabla\theta$, but it is multiplied  by a 
singular function of the phase field $\phi$, which diverges in the 
limit $\phi\to 1$ (the solid). These singular features are needed to create
stable grain boundary solutions, that is, localised spatial regions where 
the phase field departs from its solid value and the angle field exhibits
rapid variations.

Both models have a variational structure for the dynamics of the
phase field and the angle field, that is
\begin{equation}
\partial_t \phi = - M_\phi \frac{\delta{\cal F}}{\delta \phi},
\end{equation}
\begin{equation}
\partial_t \theta = - M_\theta \frac{\delta{\cal F}}{\delta \theta},
\label{thetaeq}
\end{equation}
which means that both $\phi$ and $\theta$ evolve such as to follow the
gradient of the free energy, with $M_\phi$ and $M_\theta$ being the
corresponding mobilities (which may be functions of the fields). 
In the following, it will be shown that Eq.~(\ref{thetaeq}) is 
incorrect for coherent crystalline matter.

\begin{figure}
\begin{center}
\includegraphics[width=\textwidth]{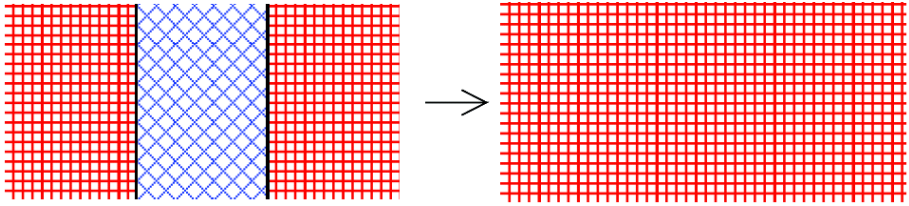}
\caption{Evolution of a tricrystal in the orientation-field models.
The crystalline slab in the centre rotates, and eventually the grain
boundaries disappear.}
\label{fig1}
\end{center}
\end{figure}

To illustrate the problems with this equation of motion, it is again
useful to analyse a simple one-dimensional situation, which is a
tricrystal. A slab of crystalline orientation $\theta_0$ is sandwiched
between two crystals of identical orientation $\theta=0$, as shown in
the left side of Fig.~\ref{fig1}. The two crystals on the sides of the system
are assumed to be clamped to a substrate, that is, $\theta=0$ for all
times. In both models, this initial condition evolves with time: the
orientation of the central slab remains homogeneous, but changes with
time to approach the orientation of the outer crystals. The final 
state is a uniform solid of orientation $\theta=0$: the
central slab has disappeared. 

Of course, this process can take place since it corresponds to
a minimisation of the free energy: the two grain boundaries with 
their positive grain boundary energy are eliminated. However, the 
pathway of this dynamics is not appropriate for the evolution of 
a coherent crystal. In fact, Eq.~(\ref{thetaeq}) corresponds
to the dynamics of matter which has orientational, but no
positional order, such as a liquid crystal. Indeed, if in model
I the term proportional to $|\vec\nabla\theta|$ is omitted or
in model II the singular coupling function is replaced by a 
regular one, the resulting model can be mapped to the standard 
Landau-de Gennes model for nematic liquid crystals in two 
dimensions \cite{deGennes}. The free energies in Eqs.~(\ref{modelI}) 
and (\ref{modelII}) have been designed to stabilise grain 
boundaries, which do not exist in a nematic liquid crystal. 
The energetics of the models are thus quite different from liquid
crystals. In contrast, the type of the dynamics has stayed the same. 

To understand where is the difference in dynamics between liquid
crystals and crystals, consider the elongated 
molecules of a nematic liquid crystal characterised by a 
director field of a certain orientation $\theta_0$. Since the 
molecules have no bonds, it is possible to change the local 
orientation while keeping the centres of mass fixed, by just
making each molecule rotate around its centre of mass (of course,
in a dense liquid crystal, this exact procedure is not possible 
because of steric exclusion, but the director can still be
changed with only short-range displacements of the centres of
the molecules). The system is thus free to {\em locally} change
orientation in order to lower its free energy, and thus 
follows Eq.~(\ref{thetaeq}). This is obviously not the 
case in crystalline matter: it is not possible to rotate a unit 
cell without displacing the surrounding neighbours, because bonds 
(or, more generally, the positional ordering of elements) define 
a connectivity. It is easy to grasp that the evolution depicted
in Fig.\ref{fig1} is impossible if the connectivity of the central 
slab is preserved.

Thus, a consistent evolution equation for $\theta$ has to take 
into account this connectivity, or, in other words, the evolution
of the positions. This is, in general, a complicated undertaking.
Two elementary situations where it easy to obtain an equation are 
(i) rigid body rotation, in which case the (advected) time derivative
of the local angle is given by the curl of the local velocity field,
or (ii) purely elastic deformations of the solid, in which case the
orientation is not an independent quantity but can be deduced
from the elastic displacement field.

Here, a third possibility will be briefly discussed, namely,
plastic deformation. This corresponds precisely to a change in
the connectivity of matter. If the matter in question can be
considered reasonably crystalline (as opposed to, for example,
an amorphous material), its geometry can formally always
be described by a density of dislocations, which are singularities
of the displacement field if a perfect crystal is taken
as the reference state. If, furthermore, grain boundaries
remain coherent (that is, no grain boundary sliding takes place), 
the evolution of the local orientation can be linked to the motion 
of dislocations. A complete description is far outside of the 
scope of this article; the interested reader is referred 
to Ref.~\cite{Kroener} for a detailed introduction to the
continuum theory of defects. Here, only two simple examples 
will be qualitatively treated for illustration.

\begin{figure}
\begin{center}
\includegraphics[width=.8\textwidth]{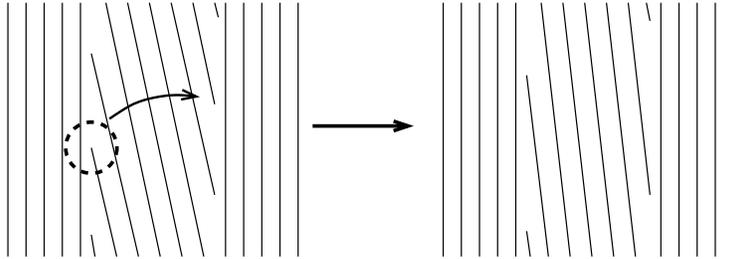}
\caption{Sketch of the elementary process that generates a rotation
of the central crystal slab by the motion of a single edge dislocation.
Only the crystal planes close to the vertical direction are shown.}
\label{fig2}
\end{center}
\end{figure}

Consider again the tricrystal configuration. In the sketch
shown in Fig. \ref{fig2}, only one set of crystal planes is shown 
for clarity, and the central slab has a small misorientation
with respect to the outer crystals. In this situation, the
two low-angle grain boundaries consist of individual edge
dislocations. The inner crystal can now rotate by an
elementary process: take one of the edge dislocations of the
left grain boundary (marked by a circle) and make it glide
towards the other grain boundary. This process involves only local 
reconnection events. When the dislocation arrives at the right 
grain boundary, it can annihilate with a dislocation of the 
opposite sign. As a result, one dislocation has disappeared from
each grain boundary. Of course, this process can repeat itself
until no dislocation is left, and the grain boundaries
have disappeared. It should be stressed that this pathway
for rearrangement exhibits large energy barriers, since the 
elastic energy of a single dislocation is much higher
in the centre of the slab than at its original position within
the grain boundary. Therefore, if only thermal fluctuations are
driving this process (no external strains), it will be extremely
slow.

On a more quantitative level, the misorientation through
a grain boundary is linked to the density of dislocations by
simple geometrical arguments. Therefore, it is natural that the 
misorientation is lowered when the dislocation density in the 
grain boundaries decreases. Furthermore, it is obvious 
that the rotation rate of the central slab is proportional 
to the current of dislocations crossing the
crystal. Thus, a consistent equation of motion for the 
orientation should be based on the
evolution of the dislocation density. However, the development
of such an equation is a difficult task, because the motion
of dislocations is determined by their complicated elastic
interactions, as well as by external strain and interactions
with other defects. Despite intense activity on the phase-field 
modelling of defects, elasticity, and plasticity (see \cite{Wang10} 
for a recent overview), such an equation seems at present out
of reach.

Let us now come back to the outcome of the simulations for the tricrystal
configuration. The functional derivative of the gradient term in
Eq.~(\ref{modelI}) of model I generates a non-local diffusion equation 
for the angle field, which has to be regularised as described in
Ref.~\cite{Warren03}. For a constant mobility, the nonlocal
interaction between the grain boundaries leads to a rotation
rate that is almost independent of the distance between the grain 
boundaries. In model II, the rotation rate of the central
crystal decreases exponentially with the distance between the
grain boundaries \cite{Henry10}. In both cases, the central
slab eventually disappears. While, quantitatively, neither
of these evolutions is likely to be accurate, qualitatively
the result is the same as the one achieved by dislocation
motion.

To see that there can be qualitative differences between the
two dynamics, consider now a circular grain of orientation $\theta_g$ 
inserted in an infinite monocrystal of orientation $\theta=0$. 
Suppose that the misorientation (which is equal to $\theta_g$) 
is small, such that the grain boundary is made of individual 
dislocations separated by a typical distance $d$ which is much 
larger than the lattice spacing. Furthermore, suppose that
the grain radius $R$ is large, $R\gg d$, such that on the scale
of the grain the boundary can still be described as a continuous 
line. For simplicity, disregard any anisotropy in the grain boundary 
energy or mobility. Then, the grain will shrink by standard motion by 
curvature, and the dislocations will simply move towards the centre of the
grain. Note that the motion of the dislocations might not be strictly
radial due to their coupling to the crystal structure; however, this 
does not change the present discussion, as long as no annihilation of
dislocations takes place. Indeed, in this case, the total number
of dislocations is conserved, and the dislocation density is simply 
proportional to $1/R$, which {\em increases} with time as the grain shrinks.
This means that the misorientation also increases with time, and
if the outer crystal is fixed, the circular inner grain has to perform 
a rigid body rotation away from the orientation of the outer crystal.
This seems surprising at first, since for low-angle grain boundaries
the grain boundary energy is an increasing function of the misorientation.
However, this process is perfectly possible if it leads to a decrease 
of the total energy of the grain boundary, which is given by 
$E_{\rm gb}=2\pi\gamma(\theta_g)R$, with $\gamma(\theta_g)$ the
misorientation-dependent grain boundary energy. Its time derivative is
\begin{equation}
\frac{dE_{\rm gb}}{dt}= 2\pi\left[\gamma(\theta_g)\frac{dR}{dt}+
  R \gamma'(\theta_g) \frac{d\theta_g}{dt}\right],
\end{equation}
where $\gamma'>0$ is the derivative of $\gamma$ with respect to the
misorientation. The evolution can thus take place if the first term,
which is always negative since $dR/dt<0$, is large enough to outweigh
the second one, which is positive. In that case, the geometrical
constraints thus predict an increase of $\theta_g$ with time. The 
orientation-field models make exactly the opposite prediction: 
since the angle field evolves {\em locally} such as to lower the 
energy, the misorientation of the inner grain should {\em decrease} 
with time. Recently, this situation was investigated by numerical
simulations \cite{Wu10} using the phase-field crystal model \cite{Elder04},
which gives a faithful microscopic picture of dislocations. An increase 
of the misorientation with time was observed, consistent with
the geometrical constraints. A previous study that had compared 
phase-field and molecular dynamics simulations \cite{Upmanyu06}
and had reached different conclusions was limited to high 
misorientations, such that the above hypotheses were not satisfied. 

In conclusion, the simple relaxation equation for the angle
field, Eq.~(\ref{thetaeq}) is not consistent with the coherent
crystalline structure of matter, and can sometimes lead to 
predictions that are even qualitatively wrong. For practical purposes, 
the quantitative importance of the committed errors might be small
when the evolution of a large-scale grain structure 
is considered, but this has to be confirmed 
for each case at hand. It is worth mentioning that 
orientation-field models have been used to investigate the
interplay between the positional and orientational degrees
of freedom during the solidification of spherulites \cite{Granasy05}
or in the presence of foreign-phase particles \cite{Granasy03NM}.
These studies were performed with a vanishing orientational
mobility $M_\theta$ in the solid, and are thus not affected by
the problem pointed out here. Indeed, in the interfacial region
where the structure of the solid in not yet fully established,
the concept of a rotational mobility is valid.

\section{Fluctuations and nucleation}
\label{sec4}

Many phase-field simulations include fluctuations, which are often 
introduced in a purely qualitative way to trigger instabilities
or to create some disorder in the geometry of the microstructures.
The role of fluctuations has been investigated more quantitatively
in connection with the formation of sidebranches in free dendritic
growth \cite{Karma99,Borzsonyi00,Bragard02}. The standard approach
is to include fluctuations as Langevin terms in the field equations,
with coefficients deduced from the fluctuation-dissipation theorem.
Before proceeding further, this procedure will be summarised.

After inclusion of noise, Eqs.~(\ref{pf}) and (\ref{diffusion}) for 
the solidification of a pure substance become (see Ref.~\cite{Karma99}
for details)
\begin{equation}
\partial_t \phi = {\vec \nabla}^2 \phi + \phi-\phi^3 -\lambda u 
(1-\phi^2)^2 + \xi(\vec r,t),
\label{pfnoise}
\end{equation}
\begin{equation}
\partial_t u = D{\vec\nabla}^2 u + \frac 12 \partial_t h(\phi)
   - \vec\nabla\cdot \vec q(\vec r,t),
\label{diffunoise}
\end{equation}
where $D(\phi)\equiv D$ is assumed (symmetric model), and lengths
and times have been scaled by the interface thickness $W$ and
the phase-field relaxation time $\tau$, respectively. Here, 
$\xi(\vec r,t)$ and $\vec q(\vec r,t)$ are random fluctuations 
of the phase field and random microscopic heat currents, respectively. 
They are assumed to be $\delta$-correlated in space and time,
\begin{equation}
\left\langle \xi(\vec r,t)\xi(\vec r',t)\right\rangle
 = 2 F_\phi \delta(\vec r-\vec r')\delta(t-t'),
\label{fnoisedef}
\end{equation}
\begin{equation}
\left\langle q_m(\vec r,t) q_n(\vec r',t)\right\rangle
 = 2 D F_u \delta_{nm} \delta(\vec r-\vec r')\delta(t-t'),
\end{equation}
with dimensionless amplitudes $F_\phi$ and $F_u$ given by
\begin{equation}
F_u = \left(\frac{d_0}{W}\right)^d F_{\rm expt},
\label{fudef}
\end{equation}
\begin{equation}
F_\phi = \frac{2\sqrt{2}}{3} \left(\frac{d_0}{W}\right)^{d-1} F_{\rm expt},
\label{fphidef}
\end{equation}
where $d$ is the spatial dimension, and the quantity 
$F_{\rm expt}$ is determined by materials properties only,
\begin{equation}
F_{\rm expt} = \frac{k_B T_m^2 c_p}{L^2 d_0^d},
\label{fexpt}
\end{equation}
where $k_B$, $T_m$, $c_p$, $L$, and $d_0$ are Boltzmann's
constant, the melting temperature, the specific heat, the latent
heat, and the capillary length, respectively. The latter is given
by $d_0=\gamma T_m c_p/L^2$, where $\gamma$ is the surface free energy.
With the help of this expression for the capillary length, $F_{\rm expt}$ 
can be rewritten as $F_{\rm expt} = k_B T_m/(\gamma d_0^{d-1})$, which
makes its physical meaning more transparent: it is the ratio of the
thermal energy and a capillary energy scale, and can thus be seen as
a non-dimensional temperature.

In a finite-difference discretization of timestep $\Delta t$ 
and grid spacing $\Delta x$,
the noise terms are implemented by drawing, at each grid point $i$
and for each time step $t$, independent Gaussian random variables of
correlation
\begin{equation}
\left\langle \xi_i^t\xi_{i'}^{t'} \right\rangle = 
\frac{2F_\phi}{(\Delta x)^d\Delta t}\delta_{ii'}\delta_{tt'},
\end{equation}
where $\delta_{ii'}$ and $\delta_{tt'}$ are now Kronecker symbols,
and similarly for $\vec q$.
This procedure was shown to yield the correct interface fluctuations
at equilibrium in numerical simulations \cite{Karma99}.

An obvious question then arises, namely, can this method also
be used to explore nucleation ? Phase-field methods
have been used recently to investigate homogeneous
and heterogeneous nucleation, both in single-phase and multi-phase
systems (see, for 
example, \cite{Granasy02,Toth07b,Siquieri07,Granasy07,Warren09}).
In particular, it was found that for high undercoolings, 
diffuse-interface models yield better agreement with experiments
than classical nucleation theory, since the size of the nuclei
is not much larger than the thickness of the diffuse interfaces;
therefore, the free energy barriers calculated in phase-field
models can differ significantly from classical nucleation theory.
Is it sufficient, then, to add thermal noise as described above 
to obtain quantitative simulations of nucleation processes ?

The answer to this question is negative. The reason is that, for
strong noise, field equations like the phase-field model are
renormalized by the fluctuations. This is a well-known fact in
statistical field theory, but its implications do not yet seem 
to have been fully appreciated in the phase-field community. 
Therefore, it is useful to briefly sketch a few calculations
that can be found in textbooks (see, for example, \cite{Parisi}).
They are, therefore, neither new nor complete; however, they 
will prepare the ground for understanding the conclusions 
on the phase-field method at the end of this
section. 

Instead of the full phase-field model, consider a
single equation for a scalar field $\phi$ that reads
\begin{equation}
\partial_t \phi = -\frac{\delta {\cal H}}{\delta \phi} + \xi(\vec r,t),
\label{langevin}
\end{equation}
where $\xi$ is a non-conserved noise that is $\delta$-correlated,
\begin{equation}
\left\langle \xi(\vec r,t)\xi(\vec r',t')\right\rangle = 
  2 T \delta(\vec r-\vec r')\delta(t-t'),
\end{equation}
with $T$ a suitably non-dimensionalized temperature (such as 
$F_{\rm expt}$, see the discussion after Eq.~(\ref{fexpt})), and the
deterministic part of the equation derives from the functional
\begin{equation}
{\cal H} = \int \left[ \frac{1}{2}(\nabla\phi)^2 + V(\phi)\right]d\vec r,
\label{hamil}
\end{equation}
where $V(\phi)$ is a local potential of the field $\phi$
(lengths, times, and energies are dimensionless). It is
important to stress that ${\cal H}$ is {\em not} a free
energy functional, but the Hamiltonian of the field theory.
Eq.~(\ref{langevin}) generates an evolution in which each 
microscopic field configuration appears with probability
\begin{equation}
P = Z^{-1}\exp(-{\cal H}/T)
\end{equation}
in the limit of infinite evolution time. Here, $Z$ is the partition function,
\begin{equation}
Z = \int \;{\cal D}\phi \exp(-{\cal H}/T),
\label{funcint}
\end{equation}
and ${\cal D}\phi$ denotes a functional integration over
the field $\phi$. The free energy is then obtained by the
standard formula ${\cal F}=-T\ln Z$.

The free energy can be calculated exactly for the case
of a quadratic potential, $V(\phi)=m^2\phi^2/2$, where $m$
is a constant.
To carry out the calculations, it is useful to consider a discrete
version of the model. For simplicity, consider as the domain
of integration $V$ a $d$-dimensional torus of size $L^d$ 
with periodic boundary conditions.
When this system is discretized with the usual finite difference
formulae using $N$ grid points in each direction and hence a
grid spacing $\Delta x=L/N$, the integral in Eq.~(\ref{hamil}) 
becomes a sum over a finite number of variables. In one dimension,
\begin{equation}
{\cal H} = \frac{1}{2}\Delta x\sum_{n=0}^{N-1}
   \left[\left(\frac{\phi_{n+1}-\phi_n}{\Delta x}\right)^2 
           + m^2\phi_n^2 \right],
\label{hamilex}
\end{equation}
with the convention that $\phi_N\equiv \phi_0$. For the discretized
system, the functional integration in Eq.~(\ref{funcint}) is replaced
by a simple integration over the field variables at each grid
point,
\begin{equation}
Z = \int \exp\left(-{\cal H}/T\right) \;\prod_{n=0}^{N-1} d\phi_n \;.
\end{equation}
Since the Hamiltonian of Eq.~(\ref{hamilex}) is a quadratic 
form in the $\phi_n$'s, this is a $N$-dimensional Gaussian 
integral which can be evaluated using standard formulae.
The most convenient way is to use a discrete Fourier transform
to find the eigenvalues of the quadratic form. The final result
for the free energy is (up to a constant that can be dropped)
\begin{equation}
{\cal F} = \frac{T}{2}\sum_{l=0}^{N-1} \ln\left(m^2 + 
  \frac{4}{(\Delta x)^2}\sin^2\frac{\pi l}{N}\right).
\end{equation}
For dimensions $d>1$, the same calculation can be repeated
without difficulties, and the result is
\begin{equation}
{\cal F} = \frac{T}{2}\sum_{l_i} \ln\left(m^2 + 
  \frac{4}{(\Delta x)^2}\sum_{i=1}^d\sin^2\frac{\pi l_i}{N}\right),
\end{equation}
where the sum is now over an independent index $l_i$ for each
dimension ($i=1\ldots d$), and is normally taken over the first
Brillouin zone, $l_i\in\{-N/2+1,N/2\}$.

For an arbitrary potential $V(\phi)$, an exact calculation is 
generally impossible. Statistical field theory has developed 
sophisticated approximation methods, in particular perturbation 
expansions. Formally, every potential
can be written as a perturbation of a quadratic potential. The 
perturbation expansion (where the expansion parameter is the
temperature, which sets the fluctuation strength) 
is cumbersome and usually visualised in 
terms of diagrams \cite{Parisi}. Fortunately, the first order 
result can be understood in a relatively simple manner if we 
are interested in homogeneous systems. More 
precisely, consider the spatial average of the field,
\begin{equation}
\bar\phi(t) = \frac{1}{L^d} \int \phi(\vec r,t) \; d\vec r,
\end{equation}
which is a fluctuating quantity. The probability distribution
of $\bar\phi$ can be written as
\begin{equation}
P(\bar \phi) \sim \exp\left(-L^d f(\bar\phi)/T\right),
\end{equation}
where $f(\bar\phi)$ is the free energy density. To first order in
the perturbation expansion,
\begin{equation}
f(\bar\phi) = V(\bar\phi) + \frac{T}{2L^d}\sum_{l_i} \ln\left(V''(\bar\phi) + 
  \frac{4}{(\Delta x)^2}\sum_{i=1}^d\sin^2\frac{\pi l_i}{N}\right),
\label{theory}
\end{equation}
where the correction to the original (``bare'') potential $V(\bar\phi)$ 
is identical to the exact result for the quadratic potential, with 
the constant $m^2$ replaced by the second derivative of the bare
potential, taken at $\bar\phi$. This results from a quadratic
approximation (second-order Taylor expansion) of the bare potential
around $\bar\phi$. The result $f(\bar\phi)$ is a renormalized 
potential for $\bar\phi$.

\begin{figure}
\begin{center}
\includegraphics[width=.53\textwidth]{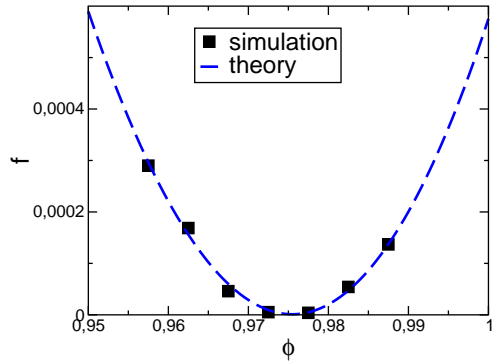}
\caption{Renormalized free energy density of the standard double-well
potential as calculated from Eq.~(\protect\ref{theory}) and from numerical
simulations, for $T=0.05$, $\Delta x=0.5$, $\Delta t = 0.005$. Only the
part close to one of the potential wells is shown. The zero of $f$ was
chosen at the minimum of the renormalized potential. The bin size for
the histograms was $\Delta\bar\phi=0.01$.}
\label{fig3}
\end{center}
\end{figure}

These calculations can be readily verified numerically. As an
example, the standard double-well potential was used,
$V(\phi)=-\phi^2/2+\phi^4/4$ (usually called $\phi^4$-potential
in the field-theory literature), and simulated in 
a two-dimensional system of size $L=32$ with a grid
spacing of $\Delta x=0.5$ and $T=0.05$, using the standard
discretization method described above with a timestep 
$\Delta t=0.005$, and an initial condition
$\phi(\vec r,0)=1$. In time intervals of $10$, $\bar\phi$ was 
calculated, and in total $1000$ points were sampled. 
Then, the free energy can be obtained by making a histogram
of the values of $\bar\phi$, and taking the logarithm of the
counts (the normalisation contributes only a constant to $f$ and 
can be disregarded). The comparison between the simulation and
the prediction of Eq.~(\ref{theory}) in Fig.~\ref{fig3} shows 
excellent agreement.

It can be seen that the minimum of the free energy density is shifted 
with respect to its ``bare'' value $\bar\phi=1$. This can be understood
intuitively by the following reasoning. The system starts in
the well of the ``bare'' potential, at $\bar\phi=1$. The random
fluctuations push the system in both directions with equal
probability, but since the potential is asymmetric, the
restoring force is larger for fluctuations towards $\bar\phi>1$
than towards $\bar\phi<1$; therefore, smaller values are more
likely to occur. In the example chosen here, the shift is small 
(the minimum is close to $1$), but for increasing temperature, the
correction becomes larger and larger (for an example of such
simulations, see \cite{Borrill97}), and eventually a phase
transition occurs (the double well disappears); in this regime,
of course the first-order perturbation result is inaccurate.

The correction also depends on the discretization. This is 
physically sound: a finer discretization introduces more degrees 
of freedom per unit volume in the discretized system, and hence allows 
for more fluctuation modes that contribute to the free energy.
With a slight change of perspective, this can also be seen as 
the natural result of a coarse-graining procedure. Indeed, if the
free energy is calculated from a given microscopic model by
coarse-graining (averaging) over cells with a certain size 
$\Delta x$ larger than the size of the microscopic elements,
both the free energy density and the amplitude of the fluctuations
that remain after the averaging (which thus have a wavelength 
larger than $\Delta x$) depend on the choice of $\Delta x$, as was
recently demonstrated explicitly for a simple lattice gas 
model \cite{Bronchart08}.

However, a problem arises in the continuum picture: it is easy to verify 
that, when the grid spacing $\Delta x$ tends to zero, the sum in
Eq.~(\ref{theory}) diverges for $d\geq 2$. This is a classical
example of an ultraviolet divergence. Thus, Eq.~(\ref{langevin})
has no continuum limit, and if it is written down in continuum
language, it is implicitly understood that an ultraviolet cutoff 
must be specified. A reasonable physical value for a cutoff in 
condensed-matter systems is the size of an atom.

Let us now discuss the implications of these facts for phase-field
modelling. Even though the above calculation have not been carried
out for the full model ($\phi$ and $u$), it is clear that renormalization 
occurs. If a phase-field model is seen as a simulation tool for
a problem that is defined in terms of macroscopic parameters, the
relevant quantities that need to be adjusted in the model are 
the renormalized ones. For instance, thermophysical 
properties are usually interpolated assuming that the phase 
field takes fixed values in the bulk phases ($\phi=\pm 1$). 
If, on average, this is no longer the case, such as in the example
of Fig.~\ref{fig3}, these interpolations become incorrect.

An obvious idea to cure this problem is to choose the ``bare''
potential such that the renormalized potential has the desired
properties. For the $\phi^4$-potential, which is renormalizable,
one may choose
\begin{equation}
V=-\frac{1+\epsilon_2}{2}\phi^2 + \frac{1+\epsilon_4}{4}\phi^4,
\end{equation}
and determine the constants $\epsilon_2$ and $\epsilon_4$
by the two conditions $f'(1)=0$ and $f''(1)=2$ using Eq.~(\ref{theory}).
For the example shown above, the values $\epsilon_2=0.0693524$ 
and $\epsilon_4=0.0208810$ indeed restore the correct 
bulk properties. However, in a quantitative phase-field model,
the macroscopic properties not only of the bulk phases, but also
of the interfaces need to be controlled. It is far from obvious 
that the above procedure, designed for homogeneous systems, will 
work. This is even more so for the critical nucleus needed to 
evaluate the nucleation barrier.

It is instructive to examine some orders of magnitude. 
In Nickel, the value of $F_{\rm expt}$ is $0.234$ \cite{Bragard02},
of order unity; it can be expected that this value is of similar
order of magnitude for other substances with microscopically 
rough interfaces. An inspection of Eqs.~(\ref{fnoisedef}--\ref{fudef})
reveals that {\em if} phase-field simulations are carried out
with the ``natural'' interface thickness, which is of the order
of the capillary length $d_0$, the fluctuations are of order
unity (recall that $F_\phi$ and $F_u$ are equivalent to $T$ in the
numerical example), and renormalization cannot be neglected.
This is a natural consequence of the fact that real solid-liquid
interfaces do indeed exhibit very strong fluctuations, as
evidenced from molecular dynamics simulations \cite{Hoyt03};
therefore, a mean-field approximation (such as the phase-field
model without noise) is not accurate.
In contrast, if (as in Refs.~\cite{Karma99,Bragard02}) a
much larger interface thickness is used, the fluctuation
strength is greatly reduced, and the difference between
``bare'' and renormalized free energy is small. Note, however,
that even in this limit a sufficient refinement of the grid
would create noticeable fluctuation corrections. We are thus
faced with the conclusion (opposite to the usual point of view 
in phase-field modelling) that the use of the simple prescription 
of Ref.~\cite{Karma99} is more precise for larger interface thickness 
and coarser grids. It is noted in passing that the concept
of the sharp-interface limit, central for the asymptotic analysis
in the deterministic case, has to be reexamined because a new
length scale (the microscopic cutoff for the fluctuations)
has been introduced.

In conclusion, it is clear that the use of the phase-field
method with fluctuations is subject to caution, at least on
small length scales. To gain a better understanding, the fluctuation
effects on the couplings of the phase-field variables need to be
investigated. Furthermore, a good control of the discretization
effects needs to be achieved; the introduction of a simple
cutoff will most likely be insufficient, since the renormalized
free energy of Eq.~(\ref{theory}) also depends on the grid 
structure. While a large body of results on these topics can
certainly be found in the field-theory literature, the development
of quantitative models for specific materials remains a
challenging task.

\section{Conclusions}

In this paper, some open questions concerning various aspects
of phase-field modelling of solidification have been discussed,
and potential future directions of research have been outlined.
The selection of topics is necessarily incomplete, both concerning 
the problems and the potential solutions. For instance, the rapid
development of the phase-field crystal approach \cite{Elder04}
and related methods currently opens up interesting new perspectives 
for the modelling of polycrystals, which are not discussed further here.

The common point of the topics treated here is that they
illustrate the dual nature of the phase-field method.
On the one hand, it is a genuine representation
of condensed-matter systems and their evolution in terms of
order parameters on a mesoscopic scale. On the other hand, 
with the help of mathematical analysis, it can be turned into
an efficient simulation tool for the solution of free boundary
problems. As in the past, the development of more efficient and
robust models for materials modelling will most likely benefit
from the pursuit and confrontation of {\em both} of these two 
complementary viewpoints. Therefore, the further development of 
the phase-field method remains an exciting research topic at the 
frontiers of physics, mathematics, and materials science.

\section*{Acknowledgements}

I thank Jean-Marc Debierre, Tristan Ducousso, Alphonse Finel, L\'aszl\'o Gr\'an\'asy, 
Herv\'e Henry, Alain Karma, Yann Le Bouar, Jesper Mellenthin, Tam\'as Pusztai, 
and James Warren for stimulating discussions on these and many 
other topics.


\label{lastpage}

\end{document}